\documentclass[lettersize,journal]{IEEEtran}

\usepackage{algorithmic}
\usepackage{array}
\usepackage{textcomp}
\usepackage{stfloats}
\usepackage{verbatim}
\usepackage{diagbox}
\usepackage{algorithm,amsbsy,amsmath,amsfonts,amssymb,epsfig,bbm,mathrsfs,multirow,amsthm,xcolor,cite}
\usepackage{epstopdf}
\usepackage{graphicx,caption,subcaption}%for three figures side-by-side
\usepackage{setspace}
\usepackage{balance}
\usepackage[hidelinks]{hyperref}
\usepackage[labelformat=simple]{subcaption}

\newcommand{\argmax}{\mathop{\rm argmax}}

\PassOptionsToPackage{hyphens}{url}
\usepackage{url}

\begin{document}

\title{Collaborative Learning for Cyberattack Detection \\ in Blockchain Networks}

\author{Tran Viet Khoa, Do Hai Son, Dinh Thai Hoang, Nguyen Linh Trung, \\Tran Thi Thuy Quynh, Diep N. Nguyen, Nguyen Viet Ha, and Eryk Dutkiewicz.	
        
\thanks{This work is the output of the ASEAN IVO \url{http://www.nict.go.jp/en/asean_ivo/index.html} project ``Cyber-Attack Detection and Information Security for Industry 4.0'' and financially supported by NICT \url{http://www.nict.go.jp/en/index.html}. This research was supported in part by the Australian Research Council under the DECRA project DE210100651. This work was supported in part by the Joint Technology and Innovation Research Centre -- a partnership between the University of Technology Sydney (UTS) and the University of Engineering and Technology, Vietnam National University, Hanoi.}

\thanks{T.~V.~Khoa is with the School of Electrical and Data Engineering, University of Technology Sydney (UTS), Sydney, Australia and the Advanced Institute of Engineering and Technology (AVITECH), University of Engineering and Technology, Vietnam National University, Hanoi, Vietnam (e-mail: khoa.v.tran@student.uts.edu.au, khoatv.uet@vnu.edu.vn).}
\thanks{D.~H.~Son is with the VNU Information Technology Institute, Hanoi, Vietnam and the Advanced Institute of Engineering and Technology (AVITECH), University of Engineering and Technology, Vietnam National University, Hanoi, Vietnam (e-mail: dohaison1998@vnu.edu.vn).}
\thanks{D.~T.~Hoang, D.~N.~Nguyen, and E.~Dutkiewicz are with the School of Electrical and Data Engineering, University of Technology Sydney (UTS), Sydney, Australia (e-mail: \{hoang.dinh, diep.nguyen, eryk.dutkiewicz\}@uts.edu.au).}
\thanks{N.~L.~Trung (corresponding author), T.~T.~T.~Quynh, and N.~V.~Ha are respectively with the Advanced Institute of Engineering and Technology (AVITECH), the Faculty of Electronics and Telecommunications and the Insitute of Artificial Intelligence, at  University of Engineering and Technology, Vietnam National University, Hanoi, Vietnam (e-mail: \{linhtrung, quynhttt, hanv\}@vnu.edu.vn).}    

}




\maketitle

\begin{abstract}
This article aims to study intrusion attacks and then develop a novel cyberattack detection framework to detect cyberattacks at the network layer (e.g., Brute Password and Flooding of Transactions) of blockchain networks. Specifically, we first design and implement a blockchain network in our laboratory. This blockchain network will serve two purposes, i.e., to generate the real traffic data (including both normal data and attack data) for our learning models and to implement real-time experiments to evaluate the performance of our proposed intrusion detection framework. To the best of our knowledge, this is the first dataset that is synthesized in a laboratory for cyberattacks in a blockchain network. We then propose a novel collaborative learning model that allows efficient deployment in the blockchain network to detect attacks. The main idea of the proposed learning model is to enable blockchain nodes to actively collect data, learn the knowledge from data using the Deep Belief Network, and then share the knowledge learned from its data with other blockchain nodes in the network. In this way, we can not only leverage the knowledge from all the nodes in the network but also do not need to gather all raw data for training at a centralized node like conventional centralized learning solutions. Such a framework can also avoid the risk of exposing local data's privacy as well as excessive network overhead/congestion. Both intensive simulations and real-time experiments clearly show that our proposed intrusion detection framework can achieve an accuracy of up to 98.6\% in detecting attacks.
\end{abstract}


\section{Introduction}\label{sec:Int}

\IEEEPARstart{B}{lockchain}~\cite{Bitcoin, ali2018applications, xie2019survey} has been emerging as a novel technology in storing and managing data with many advantages over conventional data management systems. In particular, unlike traditional centralized data management solutions, blockchain technology allows data to be stored in a distributed manner across multiple nodes. In this way, data can be accessed and processed simultaneously at multiple nodes, thus avoiding the problem of bottlenecks and single point of failure. More importantly, one of the most important features of blockchain technology is to enable data to be stored in blocks, and once a block of data is verified and placed in the chain, it cannot be modified and/or deleted. In this way, the data's integrity can be protected thanks to outstanding features of blockchain, e.g., decentralization, immutability, auditability, and fault tolerance~\cite{ali2018applications}. As a result, there are more and more applications of blockchain technology in our lives including finance, healthcare, logistics, and IoT systems~\cite{ali2018applications, xie2019survey, biswas2020interoperability, yuan2018blockchain}. 

Due to the rapid success with a wide range of applications in most areas, especially in money transfer and cryptocurrency, blockchain-based systems have been becoming targets of many new-generation cyberattacks. For example, in September 2020, KuCoin, a crypto exchange based in Singapore, announced that its system was hacked and the hackers stole over \$281 million worth of coins and tokens~\cite{Top_hack}. In May 2019, Binance, one of the biggest cryptocurrency exchange companies in the world, was reported to be hit by a major security incident. In particular, the hackers did break the exchange's security system and withdraw over 7,000 bitcoins from digital wallets, causing a total loss of approximately \$40 million for the customers~\cite{Top_hack}. Most recently, in January 2022, Chainalysis reported that North Korean hackers performed seven attacks on cryptocurrency platforms and stole nearly \$400 million from digital assets in 2021~\cite{Chainalysis}. Although most current attacks target on virtual money exchange systems, a number of blockchain applications in critical areas such as healthcare~\cite{nguyen2019proof} and food supply chains~\cite{salah2019blockchain} could be potential for attackers in the near future. If these attacks happen, they not only cause huge losses on our assets but can also lead to many serious issues related to human health and lives. Therefore, solutions to detect and prevent attacks in blockchain networks are becoming more urgent than ever.

In network security, authentication methods (e.g., two-factor authentication and biometric technology) are used to verify and identify authorized users. By implementing authentication methods, we can ensure that only authorized individuals or entities can gain access to their systems or resources. However, it is worth noting that authentication methods, despite their benefits, are inefficient in preventing attacks in blockchain networks. For example, authentication methods cannot detect and prevent Flooding of Transactions (FoT) and Brute Password (BP) attacks that are among the most common attacks in blockchain networks. In this case, intrusion detection systems can be employed to identify anomalous actions, including unauthorized commands or the execution of malicious scripts, even after a user has been authenticated. By incorporating intrusion detection systems alongside authentication methods, we can better protect the system's security against cyberattacks~\cite{bu2010distributed}. 
Several studies have been conducted to detect BP, FoT, or Man in the Middle (MitM) attacks on blockchain networks~\cite{wang2018attack, floodtran, Choi2021}. In~\cite{wang2018attack}, the authors performed BP attacks in their testbed with different devices (i.e., MacBook Pro, MacBook Air, Mobile phone, and Raspberry Pi). After that, they could detect BP attacks based on abnormally high memory and CPU consumption. In~\cite{floodtran}, the authors explored the ledger of Monero blockchain network over a period of one month. They analyzed network capacity, block size, portion of empty blocks, etc., to point out how attackers earn profit through an FoT attack. The authors in~\cite{Choi2021} proposed a method to detect MitM attacks based on blockchain technology for photovoltaic (PV) systems. They used a smart contract that stores control commands before and after transmission. They then compared these control command values to detect MitM attacks. However, these methods only focus on a specific type of attack or are applicable after attacks have already caused damage. Machine Learning (ML) has been being considered the most effective solution to detect cyberattacks in intrusion detection systems with high accuracies~\cite{chaabouni2019network, sultana2019survey, idhammad2018distributed}. The main reasons for the outstanding advantages of using ML for intrusion detection problems compared with other conventional detection methods such as signature-based and abnormally-based are threefold. First, unlike conventional intrusion detection solutions which are usually designed to detect a specific type of attack (e.g., virus, trojan, spam, and botnet), ML solutions allow to detect many types of attacks at the same time with high accuracies. For example, Deep Learning (DL) allows to detect cyberattacks in industrial automation and control systems (e.g., denial of service, reconnaissance, naive malicious response injection, and complex malicious response injection) with an accuracy of up to 97.5\%~\cite{lu2021evolutionary}. Second, while traditional solutions are often designed to detect known attacks, ML allows to detect attacks that have never been detected and reported before. For example, in~\cite{vu2020learning}, the authors showed that their DL model can detect attacks such as the DDoS attack of Mirai and BASHLITE botnets, even though such types of attacks have never been learned/trained before by this model. Last but not least, ML algorithms, especially DL, can be deployed effectively, quickly and flexibly. For example, after a deep neural network is trained, it can be deployed in different intrusion detection systems at the same time to detect cyberattacks quickly with high accuracies. In addition, when data about new types of attacks is available, we can easily update new versions of deep neural networks through transfer learning techniques~\cite{tan2018survey}.

As a result, ML has been being considered a highly-effective solution to detect cyberattacks for blockchain networks~\cite{hassan2021anomaly}. In particular, in~\cite{kumar2021distributed}, the authors proposed to use Random Forest and XGBoost to detect attacks in a blockchain-based IoT system. The results showed that this solution can identify different types of attacks and normal behaviors with an accuracy of up to 99\%. However, they only tested their results on the BoT-IoT dataset that is not real blockchain traffic. Similarly, in~\cite{alkadi2020deep}, the authors proposed an ML-based method, called bidirectional long short-term memory (BiLSTM) to detect attacks in an IoT network before the data is stored in the blockchain network. Although the results also showed that they can detect different kinds of attacks with an accuracy of up to 99\%, they were validated only on conventional network datasets such as UNSW-NB15 and BoT-IoT datasets. These datasets were collected in conventional computer networks and thus cannot reflect actual traffic in blockchain networks. In particular, these datasets have just general attacks in computer networks without specific attacks in blockchains, e.g., changes in blockchain transactions, incorrect consensus protocol, and breaking the chain of blocks. 

To the best of our knowledge, there are only few works that consider using the real blockchain traffic, e.g., try to generate artificial data or try to create data to simulate an attack for blockchain networks to train ML models such as~\cite{kim2021anomaly, liu2021lstm, cao2021blockchain}. Specifically, in~\cite{kim2021anomaly} the authors proposed a method to collect blockchain traffic data. First, they captured traffic samples from a public Bitcoin node and used them as the normal network data. Then, for the malicious traffic data, the authors performed DoS and Eclipse attacks on a target device (this device was created to become a node in the Bitcoin network). After that, the collected data was used to train an autoencoder deep learning model. This solution showed an accuracy of attack detection up to 99\%. In~\cite{liu2021lstm}, the authors used a public dataset and a private dataset from their testbed. Then, they proposed to use a Long Short-Term Memory Network (LSTM) to learn the properties of normal samples in the datasets. After that, they deployed a Condition Generative Adversarial Networks (CGAN) model to generate the artificial Low-rate Distributed DoS (LDDoS) attack samples for their blockchain dataset. The results showed the accuracy of classification up to 93\%. In addition, in~\cite{cao2021blockchain}, the authors performed a DDoS attack namely Link Flood Attack (LFA) on a simulation Ethereum network and collected the traceroute records of the network in both normal and attack behaviors. After that, the authors used the Recurrent Neural Network (RNN) to analyze the traceroute records to identify the attacks in the network. The results showed that the attack detection rate can achieve nearly 99\%. In~\cite{ramanan2021blockchain}, the authors developed a framework based on blockchain to detect only one specific attack, i.e., replay attacks, for a power system. Moreover, in~\cite{hu2019collaborative}, the authors proposed using blockchain and the Support Vector Machine (SVM) to detect cyberattacks for multimicrogrid systems. Thus, we develop a decentralized learning model that can detect different types of attacks (i.e., Denial of Service (DoS), BP, and FoT) for blockchain-based systems.

From all the above works and others in the literature, we can observe two main challenges for ML-based intrusion detection systems in blockchain networks which still have not been addressed. In particular, the first challenge is the lack of synthetic data from laboratories for training ML models. Most of the current works, e.g.,~\cite{kumar2021distributed} and~\cite{alkadi2020deep} used conventional cybersecurity datasets (e.g., UNSW-NB15 and BoT-IoT) to train data. However, these datasets were not designed for blockchain networks, and thus they are not appropriate to use in intrusion detection systems in blockchain networks. Other works, e.g.,~\cite{kim2021anomaly, cao2021blockchain, liu2021lstm}, tried to build their own datasets for blockchain networks, e.g., by obtaining the normal samples from the Bitcoin network~\cite{kim2021anomaly}, creating simulation experiment to detect the LFA~\cite{cao2021blockchain} and generating artificial attack samples by CGAN~\cite{liu2021lstm}. However, these methods have several issues. First, normal samples of transactions from the Bitcoin network may include attacks from the public blockchain network, but all collected data was classified and labeled to be normal data. Second, the simulation experiment in~\cite{cao2021blockchain} was to generate traceroute records only for the LFA so they cannot extend to other attacks. Furthermore, it is difficult to evaluate the effects of artificial attack samples in~\cite{liu2021lstm} whether they can simulate a real attack on a blockchain network or not. Another challenge we can observe here is that all of the current ML-based intrusion detection solutions for blockchain networks are based on centralized learning models, i.e., all data is collected at a centralized node for training and detection. However, this solution is not suitable to deploy in blockchains as they are decentralized networks. Specifically, nodes in blockchain networks may have different data to train, and due to privacy concerns, they may not want to share their raw data with a centralized node (or other nodes) for training processes. It is noted that the raw data means the network traffic data of a local network. This data can be classified into normal or attack data, that will be used for the learning process. It usually contains sensitive information (e.g., cryptographic keys, usage ports, or local network bandwidth) that the node does not want to share with other nodes in the network. Moreover, sending a huge amount of data to the network will not only cause excessive network traffic but also risk compromising the data integrity of blockchain networks. 

This article aims to address the aforementioned challenges by first introducing \textbf{\textit{a novel intrusion detection dataset named BNaT which stands for Blockchain Network Attack Traffic}}, created from a real blockchain network in our laboratory and then proposing an effective decentralized collaborative machine learning framework to detect intrusions in the blockchain network. Specifically, to develop BNaT, we first set up a blockchain network in our laboratory using Ethereum (an open-source blockchain software) and perform intensive experiments to generate blockchain data (including both normal and malicious traffic data). The main objectives of producing BNaT dataset are fourfold. Firstly, we collect the BNaT in a laboratory environment to have ``clean'' data samples (i.e., to ensure that the obtained data is not corrupted, error and/or irrelevant), which is especially important for training ML models. Secondly, the BNaT can be easily extended to include new kinds of blockchain attacks, e.g., 51\% or double spending attacks. Thirdly, we perform experiments with real attacks in the considered blockchain network, and thus the BNaT can reflect better the actual attack behavior of the network than simulations or by artificial attack data generated by GAN in the literature, e.g.,~\cite{liu2021lstm}. Fourthly, we collect the data in different blockchain nodes to have a complete view of the effects when the attacks are performed in a decentralized manner. After that, we develop a highly-effective collaborative learning model to make it more effective in deploying in blockchain networks to detect attacks. In particular, in our proposed learning model, working nodes in the blockchain network (e.g., fullnodes) can be used as learning nodes to collect blockchain network data (e.g., observing its incoming traffic and classifying data). Our proposed model aims to leverage knowledge learned from all the nodes in the network in a decentralized manner yet without revealing their raw data (i.e., training datasets with labels). To do so, we first design a framework in the decentralized blockchain network in which each participated learning node (i.e., fullnode in the blockchain network) deploys a deep learning model (we will explain more details in the next section) to learn from its collected data and then share its trained model with a Centralized Server (CS). The CS can be a bootnode or any fullnode in the blockchain network. The CS will then aggregate all the trained models and send the aggregated model (i.e., the global model) back to the participated learning nodes. By repeating this process, the learning nodes can gradually update their deep learning models and finally reach convergence (to the global training model). In this way, we can not only improve the accuracy of detecting cyberattacks in blockchain networks but also eliminate the risks of exposing local data of learning nodes over the network. Our proposed model can achieve an accuracy of up to 98.6\% in detecting cyberattacks in the considered network. 
Moreover, in our proposed learning model, even though the nodes do not need to share their raw data, they still can learn useful information from other nodes in the network through extracting information from shared trained models. The main contributions of this paper can be summarized as follows.

\begin{itemize} 
	\item We set up experiments in our laboratory to build a private blockchain network with the aims of not only obtaining real blockchain datasets, but also testing our proposed learning model in a real-time manner. To the best of our knowledge, this is the first dataset obtained from a laboratory for studying cyberattacks in blockchain networks, and thus we expect that our proposed BNaT dataset can promote the development of ML-based intrusion detection solutions in blockchain networks in the near future. More details of the dataset can be found at the link~\footnote{https://avitech-vnu.github.io/BNaT}.
	
	\item We build an effective tool named Blockchain Intrusion Detection (BC-ID) to collect data in the blockchain network. This tool can extract features from the collected network traffic data, filter attack samples in network traffic, and exactly label them in a real-time manner.

	\item We propose a collaborative decentralized learning model to not only improve the accuracy of identifying attacks, but also effectively deploy in decentralized blockchain networks. This model enables fullnodes in the blockchain network to effectively share their trained models to improve cyberattack detection efficiency without the need of sharing their raw data.
	
	\item We perform both intensive simulations and real-time experiments to evaluate our proposed framework. Both simulation and experimental results clearly show the outperformance of our proposed framework compared with other baseline ML methods. Furthermore, our results reveal some important information in designing and implementing learning models in blockchain networks in practice, e.g., real-time monitoring and detecting attacks.
\end{itemize} 

The rest of this paper is organized as follows. Section~\ref{sec:blockchain} provides fundamental backgrounds and our designed blockchain network together with the collaborative learning model. Section~\ref{Sec:Collaborative} presents our proposed collaborative learning model to detect cyberattacks in the blockchain network. The experiment setup, dataset collection, and evaluation method are described in Section~\ref{Sec:Framework}. After that, the experimental results and performance evaluations are discussed more details in Section~\ref{sec:results}. Finally, we conclude the paper in Section~\ref{sec:conclusion}.

\section{Blockchain Network: Fundamentals and Proposed Network Model} \label{sec:blockchain}
\subsection{Blockchain}
Blockchain is a digital ledger technology that provides a transparent, tamper-proof, and secure environment for transmitting data. This technology enables various parties to join, verify and record transactions without a trusted third party (e.g., a bank).
In a blockchain network, multiple nodes are used to simultaneously process and store data. In particular, when a node in the blockchain network receives transactions (e.g., money exchange in the Bitcoin network), it will gather all the transactions and put them in a block. This node will then start a mining process to find a ``nonce'' value for this block. It is important to note that thanks to the feature of the hash function, there is only a small set of satisfying nonce values for a block, and these values can only be found through an intensive searching process~\cite{Bitcoin}. This mining process is a special process of blockchain networks to provide proofs for validated blocks, and thus this tamper-proof can significantly enhance security for blockchain networks. After the node finds the nonce value for the mining block, this new block will be broadcast and verified by other nodes in the network. Finally, if this block is verified, it will be put into the chain (linked to the hash value of the previous block inside its header). After the block is added to the chain, it is nearly impossible to change information in this block, and thus this property can guarantee the immutability of the blockchain. 
Another aspect of blockchain is traceability due to the infeasible collision of the hash function, and thus any transaction or block can be tracked correctly. In summary, blockchain can be termed as a decentralization, immutable, traceable, and time-stamped digital data chain (ledger).
\begin{figure}[!]
    \centering
    \includegraphics[width=\linewidth]{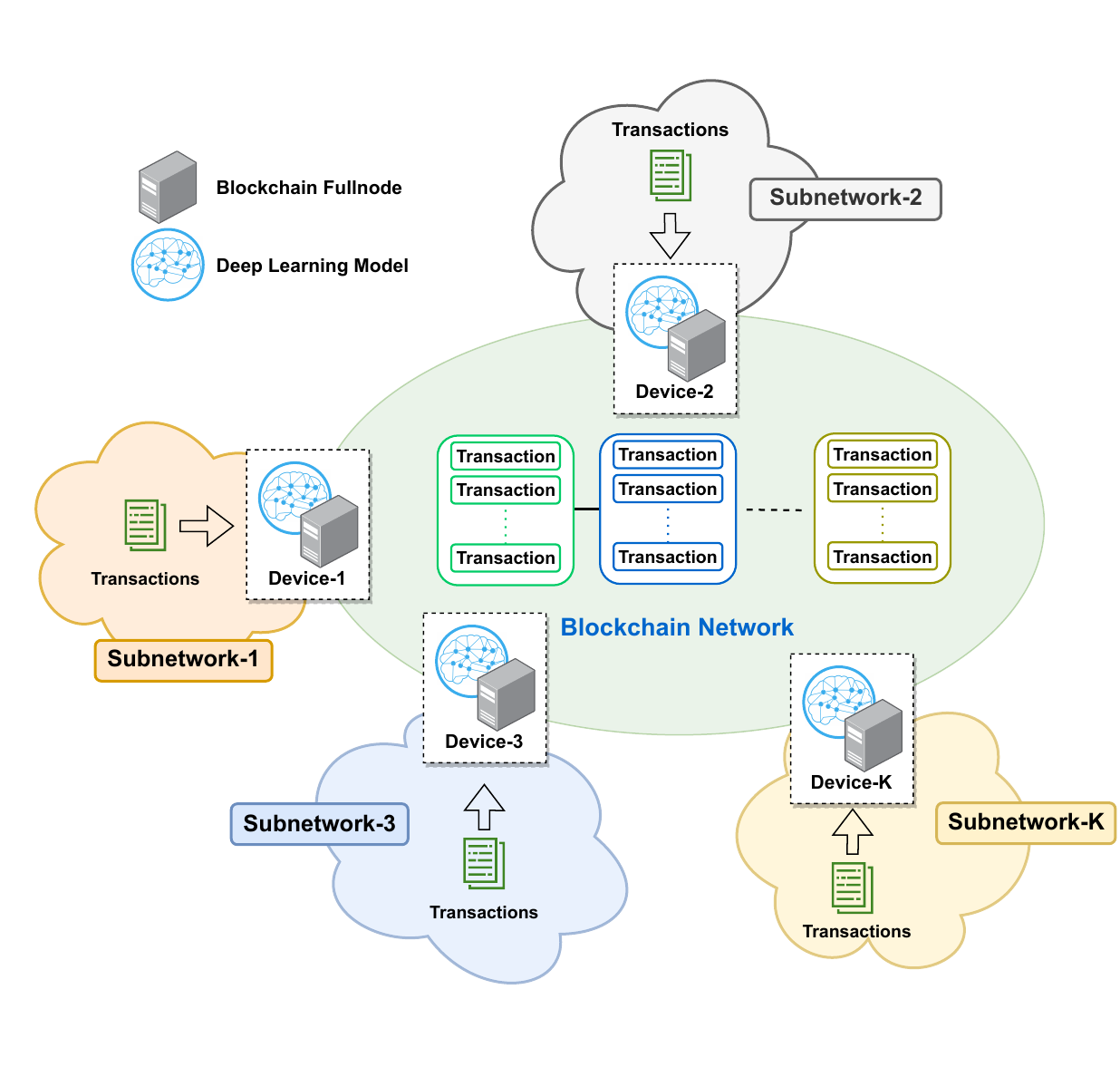}
    \caption{Our proposed learning model for blockchain network.}
    \label{fig:proposed_sys_model}
    \vspace*{-0.5cm}
\end{figure} 

\subsection{Designed Blockchain Network at our Laboratory}
To launch a blockchain network, there are two main kinds of blockchain nodes namely fullnode and bootnode. Firstly, fullnodes take responsibility to store the ledger, participate in the mining process, and verify all blocks and states. Furthermore, they can be used to serve the network and provide data on request, e.g., netstats, which is a visual interface for tracking Ethereum network status (e.g., the block number, mining status, and the number of pending transactions). Secondly, bootnode is a lightweight application used for the Node Discovery Protocol. The bootnodes do not synchronize blockchain ledger but help other Ethereum nodes discover peers to set up Peer-to-Peer (P2P) connections in the network. 

The system model together with essential components of our designed blockchain network is set up as illustrated in Fig.~\ref{fig:proposed_sys_model}. Specifically, the system includes $K$ fullnodes which are used to receive transactions, mining blocks, and keep the replica of ledger. These nodes continuously synchronize their ledgers together by the P2P protocol with equal permissions and responsibilities for processing data~\cite{Bitcoin}. In order to connect them together, a management node, known as bootnode, is set up. The fullnodes connect and interrogate this bootnode for the location of potential peers in the blockchain network. After being connected, each fullnode can collect data (i.e., transactions) from its network. Transactions can come from different blockchain applications such as cryptocurrency, smart city, food supply chain, and IoT. First, when transactions are sent to a fullnode, they will be verified and packed into one block. After the node finds the nonce value for this block, it will broadcast the block together with this nonce value to other nodes in the network for verification. Finally, if the block is verified by majority of nodes in the network, it will be added to the chain. 

At our laboratory, we design a private blockchain network based on the Ethereum blockchain network. This network also uses the Proof-of-Work (PoW) consensus mechanism, but the block confirmation time is significantly faster than the older version of Bitcoin. Furthermore, the smart contract layer of Ethereum is suitable for flexible purposes of decentralized environments as mentioned above. In addition, at each node, several attacks, that can cause serious damage to the public blockchain network, will be considered. We then capture the traffic data to analyze their impacts on the blockchain network using~\mbox{BC-ID}. Note that, in practice, there is no software that supports automatically capturing the blockchain network traffic so far. Thus, we analyze the blockchain network traffic data using a software named Wireshark~\cite{wireshark} and build a new collection tool, namely~\mbox{BC-ID} (more details will be explained in Section~\ref{Sec:Framework}). In this way, we can observe the effects of these attacks on different nodes in the blockchain network.  

\section{Proposed collaborative learning model for intrusion detection in blockchain network}\label{Sec:Collaborative}

\begin{figure}[!]
	\centering
	\includegraphics[width=\linewidth]{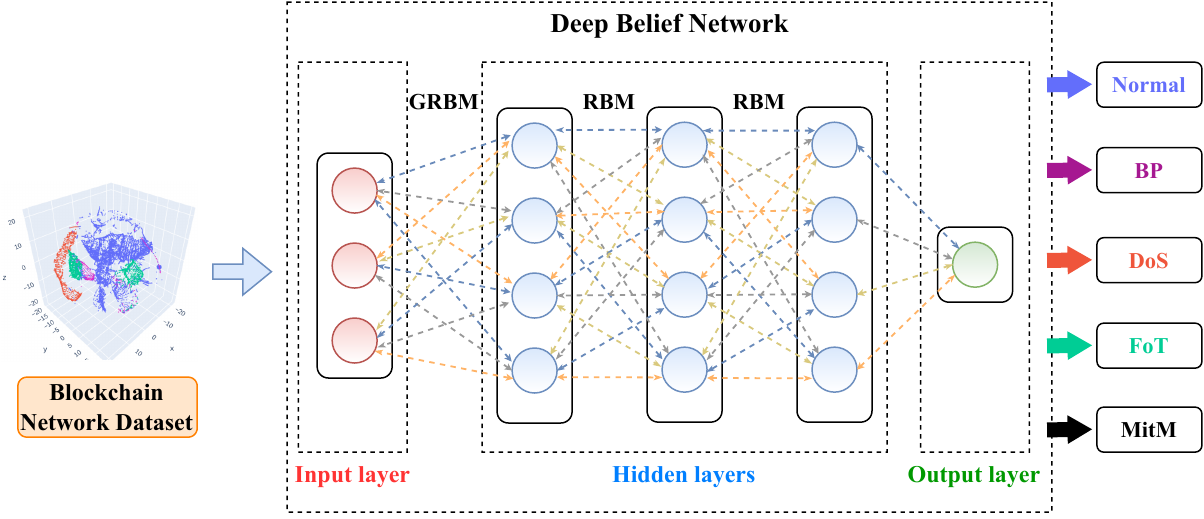}
	\caption{The structure of classification-based for intrusion detection learning model in a blockchain network.}
	\label{fig:DBN}
    \vspace*{-0.5cm}
\end{figure} 

Fig.~\ref{fig:proposed_sys_model} describes our proposed framework for intrusion detection in the blockchain network. In our proposed collaborative learning model, the fullnodes in the blockchain networks will be used as Learning Nodes (LNs) to learn knowledge from their collected data inside their subnetworks and share their learned knowledge to improve learning performance for the whole network. We also propose to use a deep neural network at each LN to learn useful information from its collected data. Then, the LNs will share their trained learning models with the CS. After that, the CS will calculate the aggregated model (i.e., the global model) and share this model back with the LNs. When a LN receives this aggregated model from the CS, it will integrate with its current LN and train its local dataset. This process will be repeated until convergence or reaching a predefined maximum number of iterations. In the end, we can obtain the global learning model for all the LNs.

In our proposed model, each blockchain node has a set of local collected data, and we propose a deep neural network (DNN) using Deep Belief Network (DBN)~\cite{DBN} to better learn knowledge from this data. The DBN is a type of deep neural network that is used as a generative model of both labeled and unlabeled data. Therefore, unlike other supervised deep neural networks which use labeled data to train the neural networks (e.g., convolutional neural networks~\cite{goodfellow}), the DBN has two stages in the training process. The first stage is the pre-training process where the DBN trains its neural network with unlabeled dataset. The second stage is the fine-tuning process where DBN uses labeled dataset to train its neural network. Thereby, the DBN can better represent the characteristics of the dataset, and thus it can classify the normal behavior and different types of attacks with high accuracies. In addition, the DBN includes multiple Restricted Boltzmann Machines (RBM) layers for latent representation~\cite{DBN}. In the DBN training process, the current layer generates latent representation by using latent representation of the previous layer as the input. Unlike other deep neural networks which also can process both labeled and unlabeled data (e.g., autoencoder deep learning network~\cite{goodfellow}), the DBN optimizes the energy function of each layer to have better latent representations of data on each RBM layer in each iteration. Thereby, the DBN is more appropriate to analyze the blockchain network traffic where the samples and features have relative coherence with each other.

The whole processes of DBN are illustrated in Fig.~\ref{fig:DBN}. Like other DNNs, the structure of DBN has three layers including an input layer, an output layer, and multiple hidden layers. As can be seen in Fig.~\ref{fig:DBN}, the Gaussian Restricted Boltzmann Machines (GRBM) layer, a type of RBM that can process real values of data, is the input layer to receive and transform the input data into binary values. We denote $k \in \{1,...,K\}$ as the number of learning nodes in the collaborative learning model, $\boldsymbol{v}^k$ and $\boldsymbol{h}^k$ to be the vectors of visible and hidden layers of LN-$k$, respectively. In addition, $M$ and $N$ are the numbers of visible and hidden neurons of GRBM. We denote $h^k_n$ and $v^k_m$ as the hidden layer-$n$ and visible layer-$m$ of LN-$k$. As defined in~\cite{RBM}, the energy function of GRBM of LN-$k$ is calculated as follows:    
\begin{equation}
\begin{aligned}
\label{eqn1}
E^k_{G}(\boldsymbol{v}^k, \boldsymbol{h}^k)&= \sum_{m=1}^M \frac{(v^k_m - b_{1,m})^2}{2\epsilon_{m}^2} - \\&\sum_{m=1}^M\sum_{n=1}^N w_{m,n} h^k_n\frac{v^k_m}{\epsilon_{m}} - \sum_{n=1}^N b_{2,n}h^k_n,
\end{aligned}
\end{equation}
where $w_{m,n}$ is the weight between visible and hidden neurons; $b_{1,m}$ and $b_{2,n}$ indicate the bias of visible and hidden neurons, respectively; and $\epsilon_{m}$ represents the standard deviation of the neuron in the visible layer. From the result of equation~(\ref{eqn1}), we can find the probability that is used in the visible layer of GRBM~\cite{RBM} as follows:
\begin{equation}
\begin{aligned}
\label{eqn2}
p^k_{G}(\boldsymbol{v}^k) = \frac{\sum_{\boldsymbol{h}^k}e^{-E_{G}(\boldsymbol{v}^k,\boldsymbol{h}^k)}}{\sum_{\boldsymbol{v}^k}\sum_{\boldsymbol{h}^k}e^{-E_{G}(\boldsymbol{v}^k, \boldsymbol{h}^k)}}.
\end{aligned}
\end{equation}

Then, we use the probability in equation~(\ref{eqn2}) to calculate the gradients of each GRBM layer with the expectation value $\Bigl \langle.\Bigr \rangle$ as follows~\cite{RBM}:
\begin{equation}
\begin{aligned}
\label{eqn3}
\nabla g^k_{G,m,n} &= \frac{\partial\log p^k_{G}(\boldsymbol{v}^k)}{\partial w_{m,n}} \\
&= \Bigl \langle{\frac{1}{\epsilon_{m}}v^k_m h^k_n\Bigr \rangle}_{dataset} - \Bigl \langle{\frac{1}{\epsilon_{m}}v^k_m h^k_n \Bigr \rangle}_{model}.
\end{aligned}
\end{equation}

Next, the gradient of GRBM layers can be calculated:
\begin{equation}
\begin{aligned}
\label{eqn4}
\nabla g^{G}_k &= \sum_{m=1}^M\sum_{n=1}^N \nabla g^k_{G,m,n} 
\end{aligned}
\end{equation}

In the next stage, we need to calculate the energy function and the gradient of RBM layers. We denote $M'$ and $N'$ are the numbers of visible and hidden neurons of RBM layers. As defined in~\cite{RBM} the energy functions of RBM layer of LN-$k$ are defined as follows:
\begin{equation}
\begin{aligned}
\label{eqn5}
E^k_{RBM}(\boldsymbol{v}^k, \boldsymbol{h}^k)&= - \sum_{m=1}^{M'} b_{1,m} v^k_m - \\&\sum_{m=1}^{M'}\sum_{n=1}^{N'} w_{m,n} v_m h^k_n - \sum_{n=1}^{N'} b_{2,n}h^k_n,
\end{aligned}
\end{equation}

The same as GRBM layers, we can calculate the gradient of each RBM layer as follows:
\begin{equation}
\begin{aligned}
\label{eqn6}
\nabla g^k_{R,m,n} = \Bigl \langle{v^k_m h^k_n \Bigr \rangle}_{dataset} - \Bigl \langle{v^k_m h^k_n \Bigr \rangle}_{model}.
\end{aligned}
\end{equation}

And the gradient of RBM layers in LN-$k$:
\begin{equation}
\begin{aligned}
\label{eqn7}
\nabla g^{R}_k &= \sum_{m=1}^{M'}\sum_{n=1}^{N'} \nabla g^k_{R,m,n} 
\end{aligned}
\end{equation}

After learning with multiple GRBM and RBM layers, we define $\boldsymbol{X}^{g,r}_k$ as the output of the last hidden layer of LN-$k$. In this paper, the output layer utilizes the softmax regression function to classify the data samples based on probability. We denote $\boldsymbol{W}^{o}$ and $\boldsymbol{b}^{o}$ as the weight matrix and bias vector between the output and the last hidden layer, respectively. We then can define the probability of the output $Z$ belonging to Class-$t$ as follows:
\begin{equation}
\begin{aligned}
\label{eqn8}
p^o_k(Z = t|\boldsymbol{X}^{g,r}_k,\boldsymbol{W}^{o},\boldsymbol{b}^{o}) &= softmax(\boldsymbol{W}^{o},\boldsymbol{b}^{o}) 
\end{aligned}
\end{equation}
where $t \in \{1,..,T\}$ is a class of the output, and $T$ refers to the total classes (including different types of attacks and normal behavior). The prediction $\boldsymbol{Z}_k$ of the probability $p^o_k$ in LN-$k$ can be calculated:
\begin{equation}
\begin{aligned}
\label{eqn9}
\boldsymbol{Z}_k= \argmax_t [p^o_k(Z = t|\boldsymbol{X}^{g,r}_k,\boldsymbol{W}^{o},\boldsymbol{b}^{o})],% \forall t \in \{1,2,\ldots,T\},
\end{aligned}
\end{equation}
where $Z$ is the output prediction. Then, we can calculate the gradient between the output layer and the last hidden layer from equation~(\ref{eqn8}) as follows:
\begin{equation}
\begin{aligned}
\label{eqn10}
\nabla g^o_k &= \frac{\partial p^o_k(Z = t|\boldsymbol{X}^{g,r}_k,\boldsymbol{W}^{o},\boldsymbol{b}^{o})}{\partial \boldsymbol{W}^{o}}.
\end{aligned}
\end{equation}

After that, the results of equation~(\ref{eqn4}), equation~(\ref{eqn7}), and equation~(\ref{eqn10}) are used to calculate the total gradient $\nabla g^{t}_k$ of DBN with multiple GRBM, RBM layers and the output layer of LN-$k$ as follows:
\begin{equation}
\begin{aligned}
\label{eqn11}
\nabla g^{t}_k &= \nabla g^{G}_k +  \nabla g^{R}_k + \nabla g^o_k.
\end{aligned}
\end{equation}

\begin{figure}[t!]
	\centering
	\includegraphics[width=.8\linewidth]{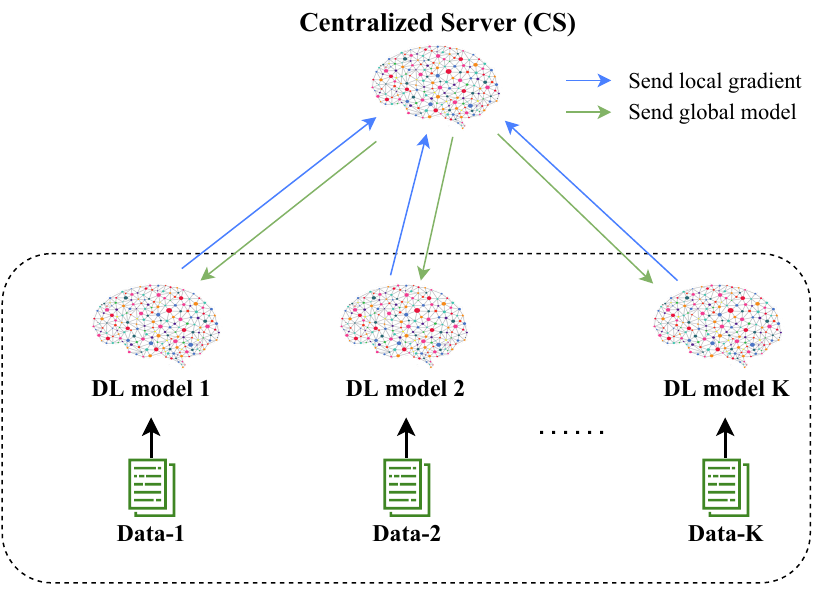}
	\caption{The illustration of the collaborative learning between DL models and the CS.}
	\label{fig:FL}
    \vspace*{-0.5cm}
\end{figure}

In the training process, the DBN first trains its neural network with unlabeled data for pre-training. Then, DBN uses its labeled data to fine-tune its neural network. At this stage, the DBN of LN-$k$ calculates its gradient $\nabla g^{t}_k$. After that, this gradient is sent to the CS to create an updated global model for all LNs as illustrated in Fig.~\ref{fig:FL}. For example, at iteration $i$ the CS receives gradients from all $K$ LNs, the CS first performs the average gradient function~\cite{Federated_learning2} as follows:
\begin{equation}
\begin{aligned}
\label{eqn12}
\nabla g^* = \frac{1}{K} \sum_{k=1}^K \nabla g_{k}^t.
\end{aligned}
\end{equation}

We then denote $\boldsymbol{\varphi}_i$ as the global model at iteration $i$ which includes the weight matrix for all layers of the LN's deep learning model, and $\mu$ represents the learning rate. From the result of equation~(\ref{eqn12}), the CS can update the global model at iteration $i+1$ as follows:
\begin{equation}
\begin{aligned}
\label{eqn13}
\boldsymbol{\varphi}_{i+1} = \boldsymbol{\varphi}_i + \mu \nabla g^*.
\end{aligned}
\end{equation}

Next, the CS sends the latest global model $\boldsymbol{\varphi}_{i+1}$ to the LNs to update their deep learning models. This process is repeated until it reaches convergence or gets the maximum number of iterations. At this time, we can find the optimal global model $\boldsymbol{\varphi}_{opt}$ that includes the optimal weights of all layers. We denote $\boldsymbol{W}^{o}_{opt}$ as the optimal weight matrix between the output layer and the last hidden layer from $\boldsymbol{\varphi}_{opt}$, the output prediction $\boldsymbol{Z}_k$ of LN-$k$ thus can be calculated as follows:
\begin{equation}
\begin{aligned}
\label{eqn14}
\boldsymbol{Z}_k= \argmax_t [p^o_k(Z = t|\boldsymbol{X}^{g,r}_k,\boldsymbol{W}^{o}_{opt},\boldsymbol{b}^{o})].% \forall t \in \{1,2,\ldots,T\},
\end{aligned}
\end{equation}

Using equation~(\ref{eqn14}), the softmax regression of each LN can classify its blockchain network samples to be a normal behavior or a type of attack. Algorithm~\ref{al:Classification_FL} summarizes the process of our proposed collaborative learning model. In our proposed model, the learning model of each network can be trained by the dataset from its local network and exchange learning knowledge with those from other nodes in a blockchain network in an offline manner. In a practical blockchain network with a large number of learning nodes, we can schedule for nodes to exchange the learning knowledge in the offline training phase at appropriate times to avoid network congestion. In this way, each node can effectively learn knowledge from other nodes while avoiding the traffic congestion of the network. After the training process, the trained models can be used to help nodes to detect attacks in a real-time manner.

\begin{algorithm}[t]
	\algsetup{linenosize=\tiny}
	\caption{The classification-based collaborative learning algorithm}
	\label{al:Classification_FL}
	\begin{algorithmic}[1]
	    \WHILE{i $\leq$ maximum number of iterations or the training process is not converged}
	        \FOR{$\forall k \in K$}
        		\STATE DBN of LN-$k$ learns $\boldsymbol{X}_k$ and produces $\boldsymbol{Z}_k$. 
        		\STATE Calculate gradient $\nabla g^t_k$.
        		\STATE Send $\nabla g^t_k$ to CS.
		    \ENDFOR
		    \STATE CS calculates average gradient $\nabla g^*$ and global model $\boldsymbol{\varphi}_i$. 
		    \STATE {$i=i+1$.}
		    \STATE CS updates global model $\boldsymbol{\varphi}_{i+1}$. 
		    \STATE CS sends global model $\boldsymbol{\varphi}_{i+1}$ to all LNs.
		    \STATE LNs update their DBNs.
		\ENDWHILE
		\STATE DBN of LNs predict and classify $\boldsymbol{Z}_k$ from the training dataset $\boldsymbol{X}_k$ with the optimal global model $\boldsymbol{\varphi}_{opt}$ .
	\end{algorithmic}
\end{algorithm}

\section{Experiment Setup, Dataset Collection and Evaluation Method}\label{Sec:Framework}

This section will explain more details about experiment setup, data collection, and feature extraction over our designed blockchain system. 

\subsection{Experiment Setup}
\label{Sec:Framework:Experiment}

\begin{figure}[t!]
	\centering
	\includegraphics[width=0.9\linewidth]{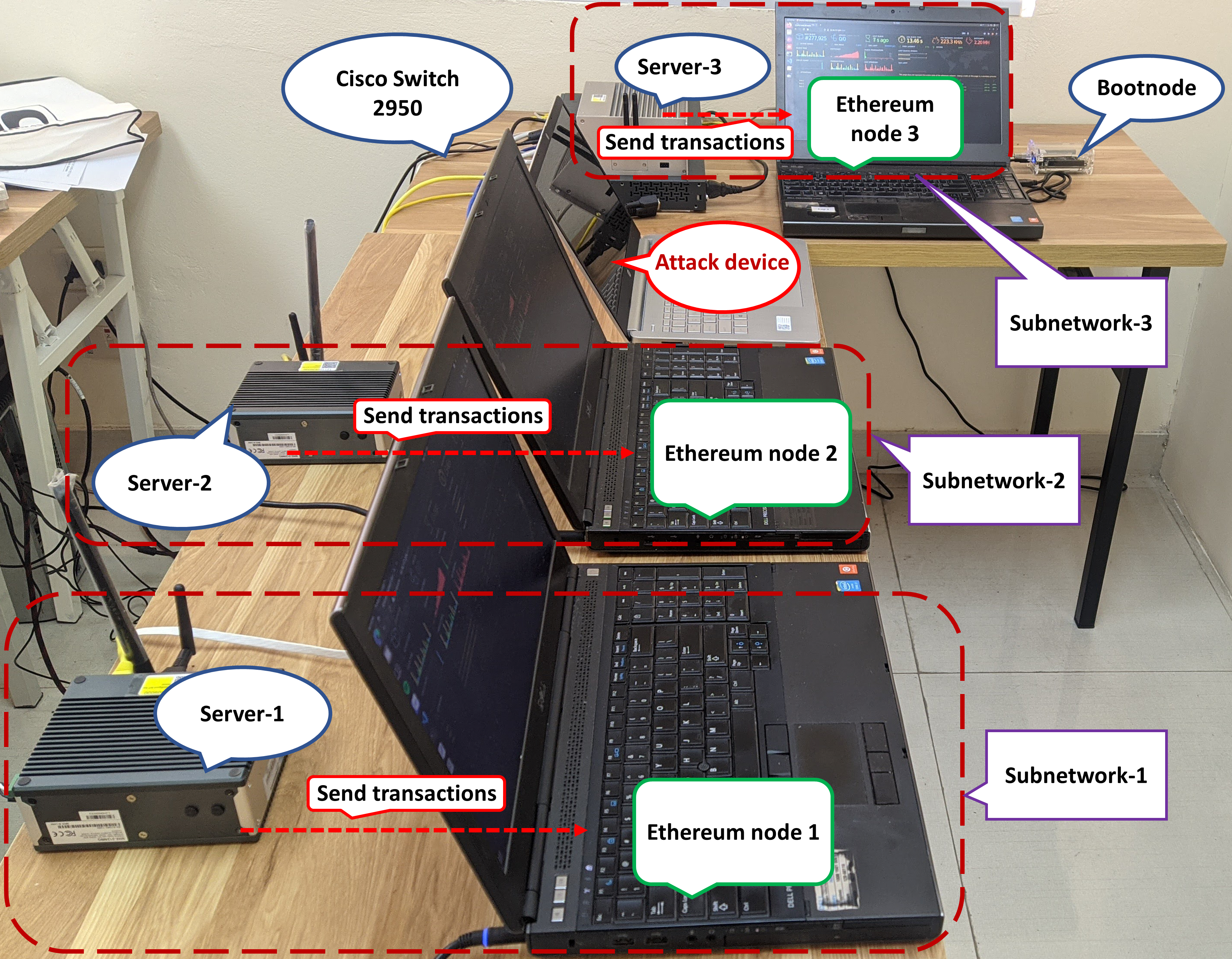}
	\caption{Experiment setup.}
	\label{fig:real_sys}
    \vspace*{-0.5cm}
\end{figure} 

In our experiments, we set up an Ethereum blockchain network in our laboratory which includes three Ethereum fullnodes, an Ethereum bootnode, and a netstats server. All these nodes are connected to a Cisco Switch Catalyst 2950 as illustrated in Fig.~\ref{fig:real_sys}. The details of these nodes are as follows:

\begin{itemize}
    \item Ethereum fullnodes are launched by \textit{Geth~v1.10.14}~\cite{Go-Ethereum} - open-source software for implementation of the Ethereum protocol. These nodes share the same initial configuration of genesis block, i.e., PoW consensus mechanism, 8,000,000 gas for block gas limit, initial difficulty 100,000. Each node is running on a personal computer with processor Intel® Core™ i7-4800MQ @2.7~GHz, RAM of 16~GB.
    
    \item Bootnode is also created by \textit{Geth~v1.10.14} and connected to the three Ethereum nodes.
    
    \item Ethereum netstats is launched by an open-source software named ``eth-netstats'' on Github~\cite{netstats}.
\end{itemize}
The normal traffic data is configured with the three trustful servers, while an attack device will execute abnormal/malicious activities to the blockchain network traffic. Each trustful server takes responsibility for generating data and sending transactions to the corresponding Ethereum node in its subnetwork as visualized in Fig.~\ref{fig:real_sys}. In summary, in the normal state, the following tasks are scheduled or randomly occur in the network:

\begin{itemize}
    \item The servers are scheduled to send transactions.
    
    \item The users call functions in the deployed smart contracts to explore the ledger. Besides transaction-related functions, the users also send requests to the Ethereum nodes for tracking their balances or the status of miners. Both of these works are randomly made by HTTP requests to the Ethereum node API (Application Programming Interface).
    
    \item Ethereum nodes broadcast transactions and mined blocks to synchronize their ledgers. The packets of bootnode are also included in this field.
    
    \item WebSockets and JSON-RPC are used when netstats get information from $Geth$ clients.
    
    \item HTTP requests and replies to view netstats interface and results of cyberattack detection.
    
\end{itemize}

\subsection{Dataset Collection and Feature Extraction}

In this section, we consider network layer aspects of the permissionless blockchain~\cite{networkattacks, networkattacks1} to detect cyberattacks in a blockchain network. In general, the goals of an adversary are usually the monetary benefit, e.g., chain splitting, and wallet theft, or stability of the network, e.g., delay and information loss. In this work, we focus on the attacks at the network layer. Attacks at the application layer, e.g., 51\%, transaction malleability attacks, timejacking, and smart contract attacks, are out of the scope of this work and can be considered in future work. Specifically, we perform four typical types of network attacks that have been reported in blockchain networks, i.e., the BP for wallet theft; DoS and MitM for information loss; and FoT for consensus delay. These are the ubiquitous attacks in the network traffic layer that have caused a number of serious consequences for many years. More details are as follows:
\begin{table}[!b]
\centering
\caption{Features of the designed dataset.}
\label{tab:features}
\resizebox{0.49\textwidth}{!}{
\begin{tabular}{|l|l|l|l|} 
\hline
\multicolumn{1}{|c|}{\begin{tabular}[c]{@{}c@{}}\#\\\end{tabular}} & \multicolumn{1}{c|}{\textbf{Features name}} & \multicolumn{1}{c|}{\textbf{T}} & \multicolumn{1}{c|}{\textbf{Description}} \\ 
\hline
\multicolumn{4}{|l|}{\textbf{Basic features}} \\ 
\hline
1 & \textit{duration} & C & \begin{tabular}[c]{@{}l@{}}length of the connection\\(seconds)\end{tabular} \\ 
\hline
2 & \textit{protocol\_type} & D & \begin{tabular}[c]{@{}l@{}}type of the protocol\\(i.e., tcp, udp, icmp)\end{tabular} \\ 
\hline
3 & \textit{service} & D & \begin{tabular}[c]{@{}l@{}}network service\\(e.g., http, ssh, etc)\end{tabular} \\ 
\hline
4 & \textit{src\_bytes} & C & \begin{tabular}[c]{@{}l@{}}number of data bytes\\from source to destination\end{tabular} \\ 
\hline
5 & \textit{dst\_bytes} & C & \begin{tabular}[c]{@{}l@{}}number of data bytes\\from destination to source\end{tabular} \\ 
\hline
6 & \textit{flag} & D & \begin{tabular}[c]{@{}l@{}}normal or error status\\of the connection\end{tabular} \\ 
\hline
\multicolumn{4}{|l|}{\textbf{Statistical features }} \\ 
\hline
\multicolumn{4}{|l|}{\textit{Features refer to source IP-based~Statistical}} \\ 
\hline
7 & \textit{count} & C & \begin{tabular}[c]{@{}l@{}}number of connections to\\the same source IP\\as the current connection\end{tabular} \\ 
\hline
8 & \textit{srv\_count} & C & \begin{tabular}[c]{@{}l@{}}number of connections to\\the same service\\as the current connection\end{tabular} \\ 
\hline
\multicolumn{4}{|l|}{\textit{Features refer to these same source IP connections}} \\ 
\hline
9 & \textit{serror\_rate} & C & \% of `SYN' errors connections\\ 
\hline
10 & \textit{same\_srv\_rate} & C & \% of same service connections \\ 
\hline
11 & \textit{diff\_srv\_rate} & C & \begin{tabular}[c]{@{}l@{}}\% of~different services\\connections~ ~\end{tabular} \\ 
\hline
\multicolumn{4}{|l|}{\textit{Features refer to these same service connections}} \\ 
\hline
12 & \textit{srv\_serror\_rate} & C & \% of `SYN' errors connections\\ 
\hline
13 & \textit{srv\_diff\_host\_rate} & C & \% of~different host connections \\ 
\hline
\multicolumn{4}{|l|}{\textit{Features refer to destination IP-based~Statistical}} \\ 
\hline
14 & \textit{dst\_host\_count} & C & \begin{tabular}[c]{@{}l@{}}number of connections to\\the same destination IP \\as the current connection\end{tabular} \\ 
\hline
15 & \textit{dst\_host\_srv\_count} & C & \begin{tabular}[c]{@{}l@{}}number of connections to\\the same service as\\the current connection\end{tabular} \\ 
\hline
\multicolumn{4}{|l|}{\textit{Features refer to these same destination IP connections}} \\ 
\hline
16 & \textit{dst\_host\_same\_srv\_rate} & C & \% of~same service connections \\ 
\hline
17 & \textit{dst\_host\_diff\_srv\_rate} & C & \begin{tabular}[c]{@{}l@{}}\% of different services\\connections\end{tabular} \\ 
\hline
18 & \textit{dst\_host\_same\_src\_port\_rate} & C & \begin{tabular}[c]{@{}l@{}}\% of same both source port\\and destination IP connections\end{tabular} \\ 
\hline
19 & \textit{dst\_host\_serror\_rate} & C & \% of `SYN' errors connections \\ 
\hline
\multicolumn{4}{|l|}{\textit{Features refer to these same service connections}} \\ 
\hline
20 & \textit{dst\_host\_srv\_diff\_host\_rate} & C & \% of different host connections \\ 
\hline
21 & \textit{dst\_host\_srv\_serror\_rate} & C & \% of `SYN' errors connections~ ~ \\
\hline
\end{tabular}
}
\end{table}

\begin{itemize}
    \item \textit{Brute Password (BP) attack}: is derived from traditional cyberattack when hackers perform such attacks to steal blockchain users' accounts. In this way, the hackers can access the users' wallets and steal their digital assets. As mentioned in Section~\ref{sec:Int}, the BP attack on KuCoin caused the loss of up to \$281 million~\cite{Top_hack}. To perform this attack, the attacker retries passwords of an Ethereum public key until it finds out the correct login information.
    
    \item \textit{Denial of Service (DoS) attack}: is also another common type of attack in blockchain networks as it can be easily performed to attack blockchain nodes. For such kind of attack, the attackers will launch a huge amount of traffic to a target blockchain node in a short period of time. Consequently, the target node will not be able to work as normal, i.e., mining transactions, and even be suspended. In the real-world, Bitfinex~\cite{Bitfinex} was temporarily suspended due to such kind of attack. Thus, in our setup, a simple DoS attack is simulated, i.e., an SYN flood attack, by repeatedly sending initial connection request (SYN) packets to an Ethereum node. 
    
    \item \textit{Flooding of Transactions (FoT) attack}: targets delay the PoW blockchain network by spamming the blockchain network with null or meaningless transactions. When the number of transactions per second in the Ethereum network suddenly hits the top, a mining node may face two following issues, i.e., too much traffic (similar as that of DoS), and the queue of pending transactions is full. It equates to the unnecessary time burden for mining process and block propagation~\cite{floodtran}. In 2017, the Bitcoin mempool size was exceeded 115,000 unconfirmed transactions which led to \$700 million worth of transaction stall~\cite{networkattacks1}. In our work, FoT attack is implemented by continuously sending a large number of transactions to an existing smart contract.

    \item \textit{Man in the Middle (MitM) attack}: is another typical attack where an attacker places himself between the legitimate communicating parties and secretly relays and possibly modifies the information exchanged between them. In this way, the attacker can intercept, read, and modify the blockchain messages. For example, hackers can read the API messages between users and blockchain nodes to steal their wallet password~\cite{wang2018attack}. To implement MitM, an attack device first filters `\textit{eth\_sendrawtransaction}' packets, which represent any transaction from users to blockchain nodes. Then, these packets' contents are randomly modified, leading to invalid transactions.
\end{itemize} 

\begin{figure*}[t]
    \centering
    \begin{subfigure}{0.22\linewidth}
        \includegraphics[width=\linewidth, height=3.6cm]{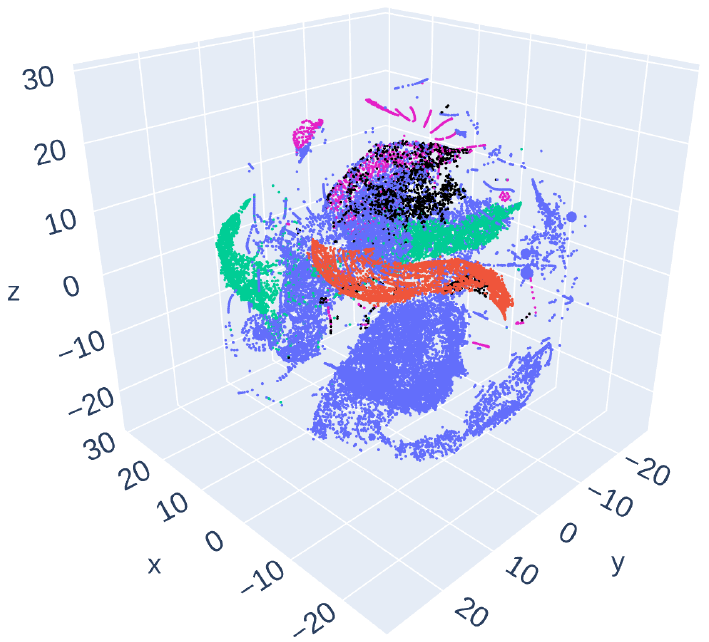}
        \caption{Learning Node 1}
        \label{fig:dataset_visu_1}
    \end{subfigure}
    \hfill
    \begin{subfigure}{0.22\linewidth}
        \includegraphics[width=\linewidth, height=3.6cm]{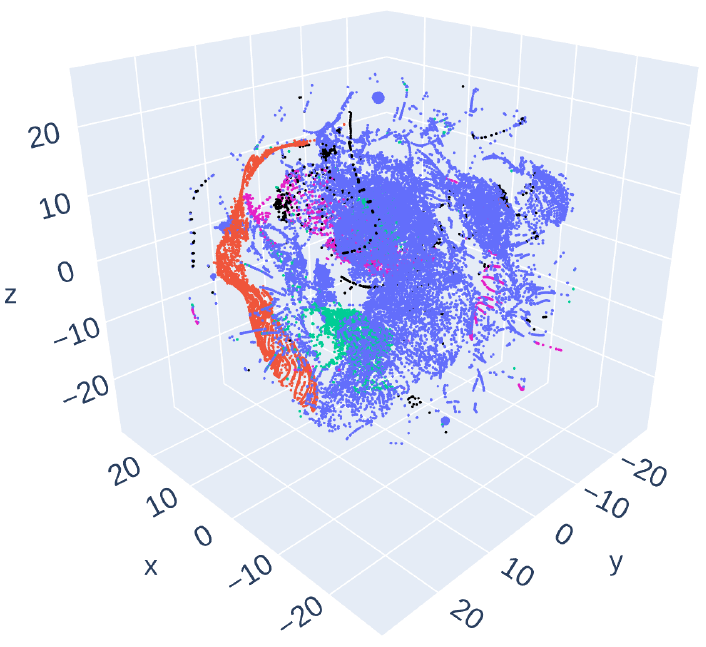}
        \caption{Learning Node 2}
        \label{fig:dataset_visu_2}
    \end{subfigure}
    \hfill
    \begin{subfigure}{0.22\linewidth}
        \includegraphics[width=\linewidth, height=3.6cm]{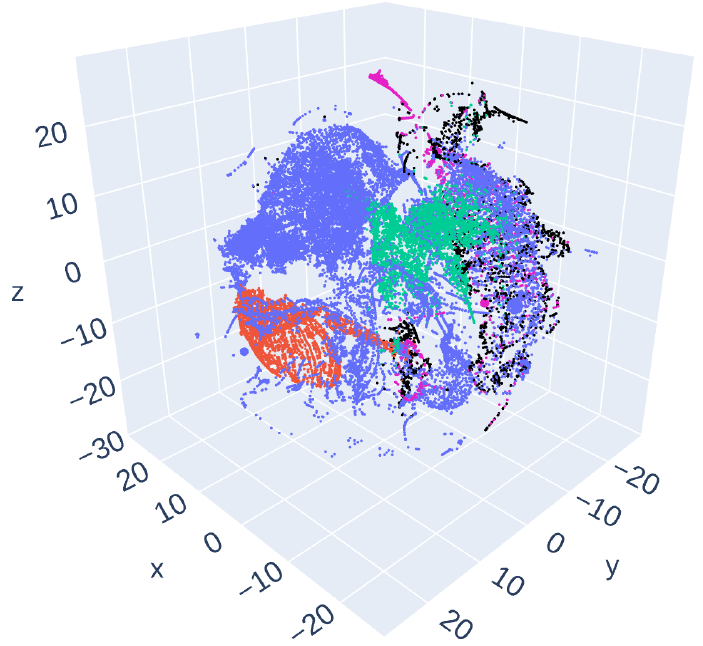}
        \caption{Learning Node 3}
        \label{fig:dataset_visu_3}
    \end{subfigure}
    \hfill
    \begin{subfigure}{0.21\linewidth}
        \includegraphics[width=\linewidth, height=3.6cm]{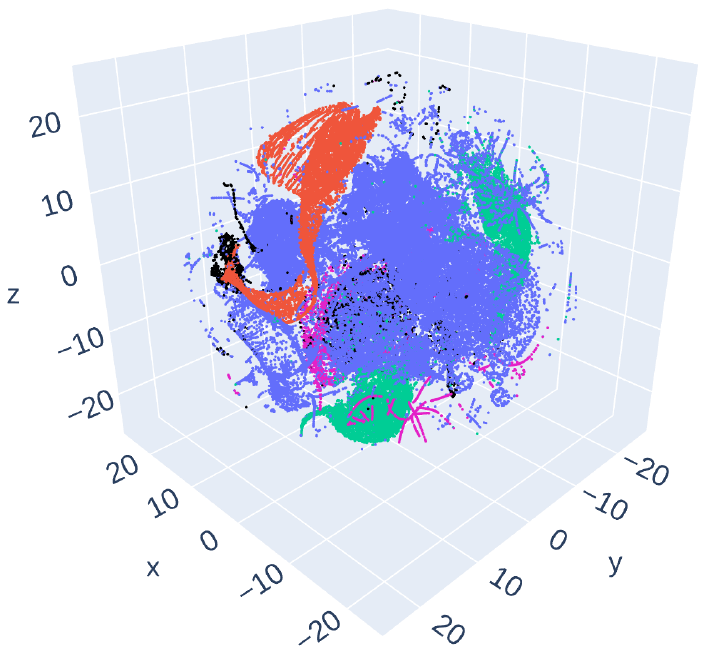}
        \caption{Combine data from all~\mbox{learning nodes}}
        \label{fig:dataset_visu_4}
    \end{subfigure}
    \hfill
    \begin{subfigure}{\linewidth}
        \centering
        \includegraphics[width=0.5\linewidth]{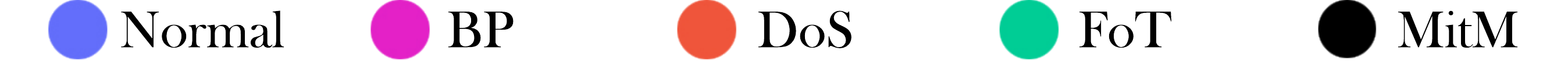}
    \end{subfigure}
    \hfill
    \caption{Visualization using \textit{t}-SNE for collected datasets.}
    \label{fig:dataset_visu}
    \vspace*{-0.5cm}
\end{figure*}

In order to capture traffic data of these attacks, we build a dataset collection tool, named~\mbox{BC-ID}, which inherits the core of an open-source utility named ``kdd99\_feature\_extractor''~\cite{kdd99fe} and our new designs to fit the considered Ethereum network, i.e., correct the service of packets related to Ethereum nodes, remove meaningless features, and automate label dataset samples based on some given properties. To do this, we first use the `libpcap-dev' library of Linux to capture all the network data (including normal and different types of attacks) from the local network. Then, the BC-ID is used to extract features from the collected data, filter the attack samples, and label them as normal or a type of attack. In particular, the BC-ID starts by capturing raw traffic data based on `libpcap-dev' package of Linux OS. Since each blockchain network has a few specific $ports$ for peer connections, client connections, and so on, BC-ID targets to filter and analyze traffic data in these ports. For example, the Ethereum blockchain network uses port 30303 for the TCP port listener, and port 8545 for JSON-RPC by default. As KDD99 dataset~\cite{kdd99}, BC-ID extracts features and then separates them into two categories, e.g, basic features (i.e., all the attributes can be extracted from a TCP/IP connection) and traffic features (i.e., statistics of packets with the same destination host or service in a window interval). 
Especially, our goal is to achieve a trained model that can be applied to our proposed real-time blockchain attack detection system, when the number of samples in a prediction frame is limited. Thus, the~\mbox{BC-ID} collects the dataset frames in which each frame lasts for 2~seconds and extracts their features. The~\mbox{BC-ID} then puts all collected data in this frame into a single file. Finally, we merge all single files together to make the full dataset. In summary, Table~\ref{tab:features} shows 21~features in the designed dataset, which are separated into two types, i.e., discrete (D) and continuous (C). For continuous features, they are calculated in 2~seconds time window (similar to that of the famous KDD99 dataset~\cite{kdd99}). 

In each Ethereum node, the separated dataset is collected in five states (classes), i.e., normal state~(Class-0), BP attack~(Class-1), DoS attack~(Class-2), FoT attack~(Class-3), and MitM attack~(Class-4). The normal state is captured in two hours, the rest of them in an hour through the designed~\mbox{BC-ID} tool. As described above, when a node is attacked, the normal traffic still exists. Therefore, the attack samples can be filtered out by features-based two properties, i.e., the source and destination IP address of the attack~device; \textit{service}, \textit{src\_length}, and \textit{dst\_length} of the samples, which are analyzed by Wireshark~\cite{wireshark}. To improve the diversity of the designed dataset, the normal traffic data in the attack state is mixed with traffic data in the normal state. In our experiments, a number of random samples in each state are selected to reduce the size of the bulk dataset as shown in Table~\ref{tab:dataset}. In fact, we mix normal traffic data in an equal ratio, i.e., 10,000 samples per normal state, normal traffic data at BP, DoS, FoT, and MitM, respectively. 

\begin{table}[!t]
    \centering
    \caption{The number of samples in the designed dataset.}
    \label{tab:dataset}
    \begin{tabular}{|l|c|c|c|}
    \hline
    \multicolumn{1}{|c|}{\textbf{\diagbox[width=\dimexpr \textwidth/6-\tabcolsep\relax, height=1.2cm]{Class}{Ethereum \\ node}}} &
      \textbf{\begin{tabular}[c]{@{}c@{}}Node-1\\ (samples)\end{tabular}} &
      \textbf{\begin{tabular}[c]{@{}c@{}}Node-2\\ (samples)\end{tabular}} &
      \textbf{\begin{tabular}[c]{@{}c@{}}Node-3\\ (samples)\end{tabular}} \\ \hline
    Normal                                                             & 50,000 & 50,000 & 50,000 \\ \hline
    BP                                                     & 5,000  & 5,000  & 5,000  \\ \hline
    DoS                                                                & 5,000  & 5,000  & 5,000  \\ \hline
    FoT & 5,000  & 5,000  & 5,000  \\ \hline
    MitM & 5,000  & 5,000  & 5,000  \\ \hline
    \end{tabular}
    \vspace*{-0.5cm}
\end{table}

Fig.~\ref{fig:dataset_visu} illustrates the visualization of our designed dataset using~\mbox{\textit{t}-Distributed Stochastic Neighbor Embedding}~\mbox{(\textit{t}-SNE)~\cite{tsne}} with three most important components.
Although sharing the same configurations for \textit{t}-SNE, the dataset of each LN has a different distribution in the output.
In 3D view, the DoS and FoT attack samples show a fairly clear separation from normal state points. Otherwise, the BP and MitM attack samples collide with the normal state samples. This indicates that discriminating BP and MitM samples from the normal data points would be significantly challenging. 

\subsection{Evaluation Method}

The confusion matrix with accuracy, precision, and recall proposed in~\cite{confusion_matrix1} is widely used to evaluate the performance of machine learning algorithms. Let TP, FP, TN, and FN denote ``True Positive'', ``False Positive'', ``True Negative'', and ``False Negative'', respectively. The accuracy of the total system with $T$ classes including normal behaviors and different types of attacks is as follows:
\begin{equation}
\begin{aligned}
\label{eqn15}
	Accuracy=\frac{1}{T}\sum_{t=1}^{T} \frac{\mbox{TP}_t+\mbox{TN}_t}{\mbox{TP}_t+\mbox{TN}_t+\mbox{FP}_t+\mbox{FN}_t}.
\end{aligned}
\end{equation}

The precision of class $t$ is calculated as $P_{re}^t = \frac{TP_t}{TP_t+FP_t}$. In this paper, we use weighted average precision to evaluate the performance of the whole system. We denote $S_t$ as the number of samples of class $t$ and $S$ as the number of samples of the whole dataset. The weighted average precision is calculated as follows:
\begin{equation}
\begin{aligned}
\label{eqn16}
	Precision= \sum_{t=1}^{T} \frac{ P_{re}^t \times S_t}{S}.
\end{aligned}
\end{equation}

The recall of class $t$ is calculated by $R_e^t = \frac{TP_t}{TP_t+FN_t}$. The weighted average recall that we use to calculate the performance of the total system is calculated as follows:
\begin{equation}
\begin{aligned}
\label{eqn17}
	Recall = \sum_{t=1}^{T} \frac{ R_e^t \times S_t}{S}.
\end{aligned}
\end{equation}

In the next section, we also use accuracy, precision, recall to evaluate and compare the performance of our proposed Collaborative Learning model (proposed CoL) with two other baseline methods, i.e., Centralized Learning model (CeL) and Independent Learning model (IL).

\section{Experimental Results and Performance Evaluation}\label{sec:results}

In this section, we use the collected datasets of three nodes described in the aforementioned section for the corresponding LNs. The dataset of each LN is randomly split into training and testing dataset. All LNs use DBN with the same structure of neural network for learning and detecting attacks. However, the LNs have to work in different learning models and various scenarios. Each LN has itself training and testing dataset, and thus we can use these datasets to evaluate and compare the performance of the proposed CoL, the CeL and the IL in different scenarios.     

\subsection{Simulation Results}

\begin{table*}[t]
\centering
\caption{Simulation results.}
\label{tab:simu_result}
\begin{tabular}{|l|c|c|c|c|c|c|c|c|c|} 
    \hline
    \multirow{3}{*}{\textbf{\diagbox[width=\dimexpr \textwidth/8-\tabcolsep\relax, height=1.25cm]{}{Model}}} & \multicolumn{4}{c|}{\textbf{2 Learning Nodes (LNs)}} & \multicolumn{5}{c|}{\textbf{3 Learning Nodes (LNs)}} \\ 
    \cline{2-10}
     & \multirow{2}{*}{\textbf{Proposed CoL~}} & \multirow{2}{*}{\textbf{CeL}} & \multicolumn{2}{c|}{\textbf{IL}} & \multirow{2}{*}{\textbf{Proposed CoL~}} & \multirow{2}{*}{\textbf{CeL}} & \multicolumn{3}{c|}{\textbf{IL}} \\ 
    \cline{4-5}\cline{8-10}
     &  &  & \textbf{LN-1} & \textbf{LN-2} &  &  & \textbf{LN-1} & \textbf{LN-2} & \textbf{LN-3} \\ 
    \hline
    \textbf{Accuracy} & \textbf{97.427} & 97.330 & 97.036 & 96.865 & \textbf{97.276} & 97.270 & 96.827 & 96.731 & 96.825 \\ \hline
    \textbf{Precision} & \textbf{93.861} & 93.620 & 92.793 & 92.000 & \textbf{93.448} & 93.249 & 92.209 & 91.343 & 92.798 \\ \hline
    \textbf{Recall} & \textbf{93.567} & 93.324 & 92.590 & 92.162 & \textbf{93.189} & 93.176 & 92.067 & 91.829 & 92.063 \\ \hline
\end{tabular}
\end{table*}

In this section, we present the simulation results with the dataset of LNs in different learning models. The details of datasets using for simulation are as follows:

\begin{itemize}
    \item \textbf{Proposed Collaborative Learning Model (proposed CoL)}: Each LN learns its training dataset and performs collaborative learning with other LNs to generate the global model. Then, we use the global model to test the merged testing dataset of all participated LNs. 
    
    \item \textbf{Centralized Learning Model (CeL)}: The centralized node (e.g., one of the mining node in the network) is assumed to be able to collect data from all the nodes in the network and train the deep learning model on the collected datasets. Then, we use the trained model to test data based on the merged testing dataset of all participated LNs. 
    
    \item \textbf{Independent Learning Model (IL)}: Each LN learns its training dataset without sharing knowledge with other LNs. Then, we use this model to test data based on the merged testing dataset of all participated LNs.
\end{itemize}

\subsubsection{Convergence Analysis}

\begin{figure}[t!]
	\centering
	\includegraphics[width=0.9\linewidth]{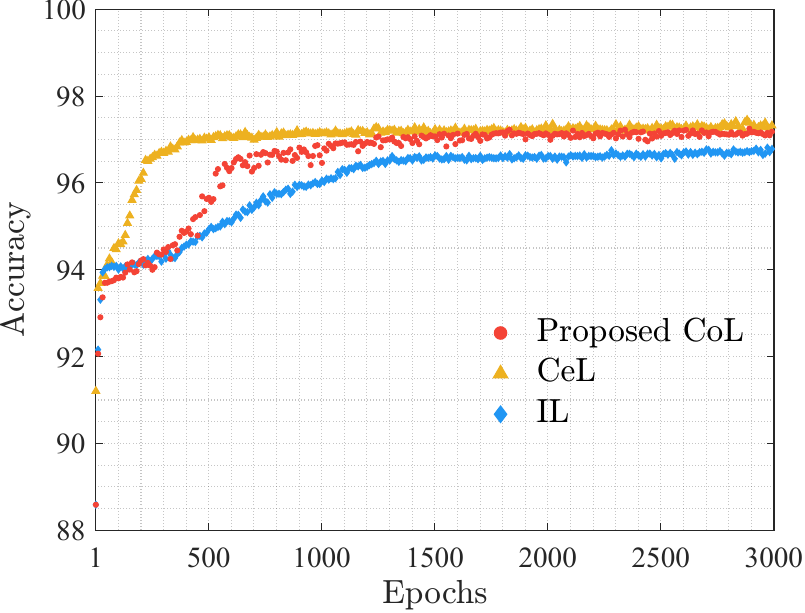}
	\caption{Training process of considered learning models.}
	\label{fig:training}
    \vspace*{-0.5cm}
\end{figure}

Fig.~\ref{fig:training} describes the convergence of the proposed CoL, the CeL and the IL (in terms of accuracy) of three LNs in the training process. The proposed CoL is obtained at the LN-1 after obtaining the global model. The CeL has a large number of training samples from three LNs so it can reach the convergence with around 97\% accuracy after 400 epochs. Besides, the proposed CoL and IL converge after 800 epochs and 1300 epochs, respectively. After 3,000 learning epochs, we can observe that the proposed CoL has a higher accuracy compared with that of the IL (i.e., 97.2\% vs 96.8\%). The reason is that the proposed CoL can obtain the exchange knowledge from DL models of other LNs. Thereby, it can achieve similar performance as that of the CeL. 

\subsubsection{Performance Analysis}

Table~\ref{tab:simu_result} presents the simulation results in two cases, i.e., two participated LNs, and three participated LNs. In both cases, we can observe the same trend when the accuracy, precision and recall of the proposed CoL are higher than those of the IL and nearly equal to those of the CeL. In particular, the accuracy of the proposed CoL is higher than that obtained by LN-1 in IL (approximately 0.5\%), and the precision of the proposed CoL is about 2\% higher than that obtained by LN-2 in IL in the case of three participated LNs. 
These results demonstrate that the proposed CoL can exchange knowledge with other LNs to improve its ability of detection, so it can achieve better performance in classifying attacks in the blockchain network than those of the IL. It also demonstrates that the learning model of IL should not be used to classify the data of other LNs. In addition, without sharing LN's dataset with a central node for training (e.g., a cloud server), the proposed CoL can achieve nearly the same accuracy as those of the CeL in all the scenarios. 

\subsection{Experimental Results}

\begin{table*}[t]
\centering
\caption{Real-time experimental results of 3 LNs models.}
\label{tab:real_result}
\begin{tabular}{|l|c|c|c|c|c|c|c|c|c|c|} 
\hline
\multirow{3}{*}{\textbf{\diagbox[width=\dimexpr \textwidth/8-\tabcolsep\relax, height=1.25cm]{}{Model}}} & \multicolumn{4}{c|}{\textbf{2 Learning Nodes (LNs)}} & \multicolumn{6}{c|}{\textbf{3 Learning Nodes (LNs)}} \\ 
\cline{2-11}
 & \multicolumn{2}{c|}{\textbf{Proposed CoL}} & \multicolumn{2}{c|}{\textbf{CeL }} & \multicolumn{3}{c|}{\textbf{Proposed CoL}} & \multicolumn{3}{c|}{\textbf{CeL }} \\ 
\cline{2-11}
 & \textbf{LN-1} & \textbf{LN-2} & \textbf{LN-1} & \textbf{LN-2} & \textbf{LN-1} & \textbf{LN-2} & \textbf{LN-3} & \textbf{LN-1} & \textbf{LN-2} & \textbf{LN-3} \\ 
\hline
\textbf{Accuracy} & \textbf{98.611} & \textbf{98.242} & 98.464 & 98.097 & \textbf{98.440} & \textbf{98.131} & \textbf{97.686} & 98.503 & 98.192 & 97.771 
\\ 
\hline
\textbf{Precision} & \textbf{97.433} & \textbf{96.871} & 97.146 & 96.634 & \textbf{97.146} & \textbf{96.717} & \textbf{95.902} & 97.138 & 96.679 & 95.864 \\ 
\hline
\textbf{Recall} & \textbf{96.529} & \textbf{95.606} & 96.159 & 95.243 & \textbf{96.101} & \textbf{95.328} & \textbf{94.214} & 96.258 & 95.481 & 94.427\\
\hline
\end{tabular}
\end{table*}

In this section, we present the experimental results obtained through real-time experiments at our laboratory. In this experiment, each blockchain node is installed a learning model to become an LN. Each LN learns its local dataset and then performs real-time attack detection in the blockchain network. We consider the scenario of two LNs and three LNs with the proposed CoL and the CeL. In the training process, the proposed CoL and the CeL are fed with the similar datasets as explained in the previous section. We then implement the trained model to all the participated LNs to perform real-time attack detection for both learning models in the testing process. 

\subsubsection{Real-time capturing and processing}
\label{Sec:Framework:Verification}

\begin{figure}[t!]
	\centering
	\includegraphics[width=\linewidth]{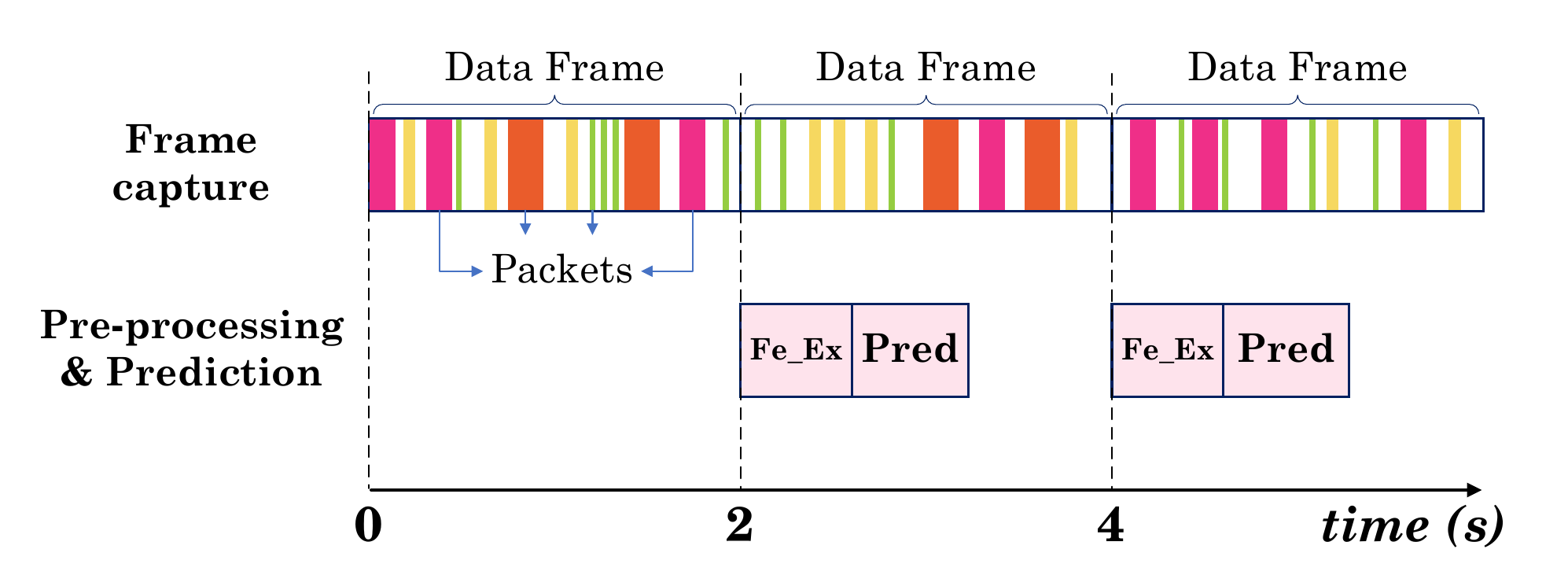}
	\caption{Timeline of verification phase.}
	\label{fig:Real_exp_timeline}
    \vspace*{-0.5cm}
\end{figure} 

In a real-time system, the cyberattack detection system continuously receives a number of the Ethereum network traffic data. Therefore, the system has to perform capturing, collecting frames, extracting features, analyzing and predicting within a period of time, i.e,~2~seconds. Fig.~\ref{fig:Real_exp_timeline} shows the timeline of the cyberattack detection program. The data frame is exploited by our feature extractor function (Fe\_Ex) of~\mbox{BC-ID} tool, and this is input for the trained model to predict (Pred) and classify packets to be normal or attack. All processes have to complete in~2~seconds before the next data frame of IP packets coming. To verify the predicted results from the trained model, all frames and prediction results are stored. These frames are merged into a full validation dataset and labeled by own designed~\mbox{BC-ID}. After that, these ground truth labels are compared with the prediction results to obtain a validation report.

\subsubsection{Performance Analysis}

Table~\ref{tab:real_result} presents the experimental results of the proposed CoL and the CeL with different participated LNs. We obtain the same trends as those of the simulation results. The accuracy results obtained by two and three learning models of both proposed CoL and CeL are slightly higher than those of the simulations at about 1\%. This is because each type of attack has different distributions of attack samples in a period of time. Table~\ref{tab:realsample} presents the number of samples of each class collected in 15 minutes. In this table, we can observe that Class-1 and Class-4 have small numbers of samples during this period, this can lead to low accuracy in statistics for these classes and reduce the total accuracy of the model. However, our proposed CoL still has better performance than those of the CeL in LN-1 in the case of two LNs (i.e., up to 0.2\% accuracy, 0.3\% precision and 0.4\% recall). Overall, our proposed CoL always achieves the best performance with approximately 98.6\% accuracy, 95.43\% precision and 96.52\% recall with two LNs and 98.44\% accuracy, 97.14\% precision and 96.1\% recall with three LNs. These results demonstrate that our proposed CoL can detect attacks with nearly the same accuracies for all participated LNs as those of CeL.

\subsubsection{Real-time Monitoring and Detection}

\begin{figure*}[t]
    \centering
    \begin{subfigure}{0.3\linewidth}
        \includegraphics[width=\linewidth]{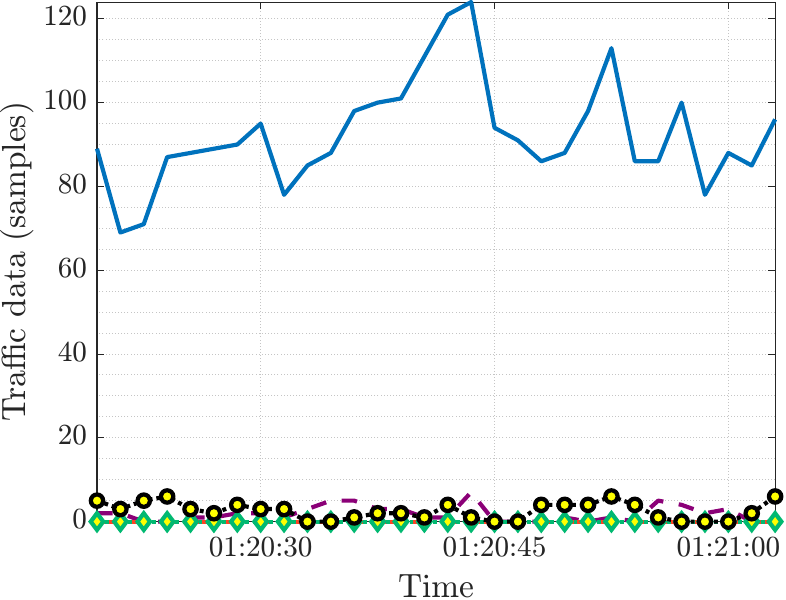}
        \caption{Normal state}
        \label{fig:realdec_1}
    \end{subfigure}
    \hfill
    \begin{subfigure}{0.3\linewidth}
        \includegraphics[width=\linewidth]{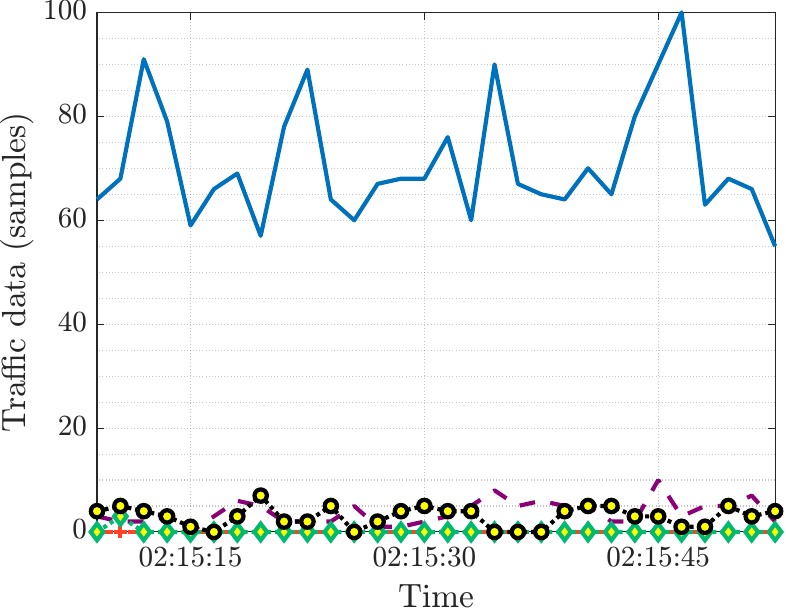}
        \caption{BP attack state}
        \label{fig:realdec_2}
    \end{subfigure}
    \hfill
    \begin{subfigure}{0.3\linewidth}
        \includegraphics[width=\linewidth]{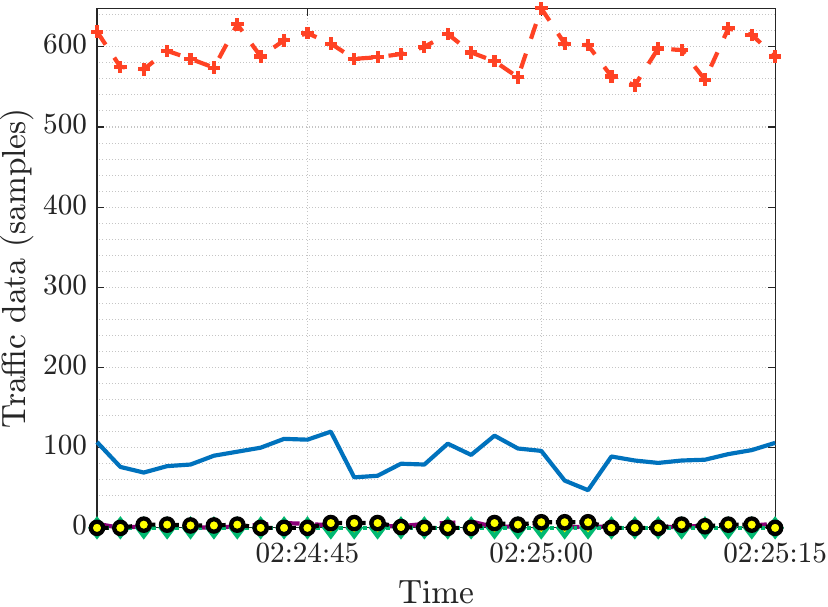}
        \caption{DoS attack state}
        \label{fig:realdec_3}
    \end{subfigure}
    \hfill
    \begin{subfigure}{0.3\linewidth}
        \includegraphics[width=\linewidth]{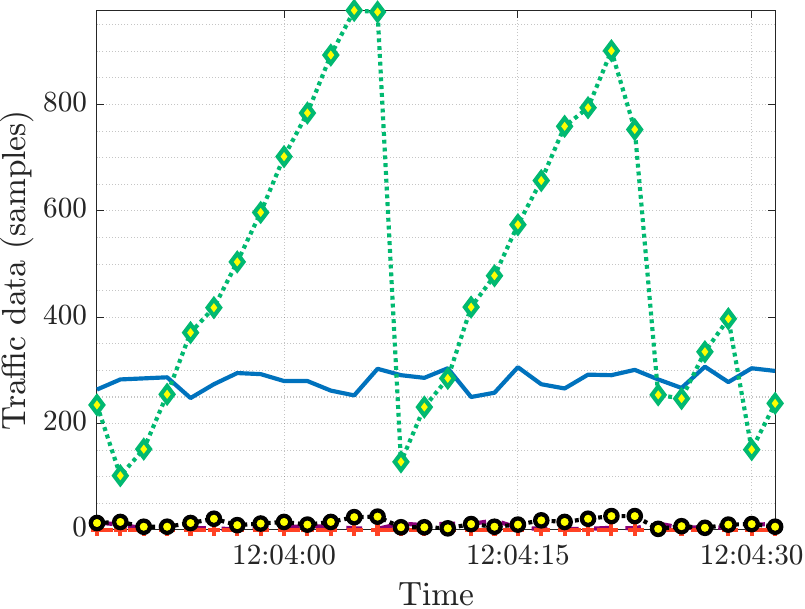}
        \caption{FoT attack state}
        \label{fig:realdec_4}
    \end{subfigure}
    \hfill
    \begin{subfigure}{0.3\linewidth}
        \includegraphics[width=\linewidth]{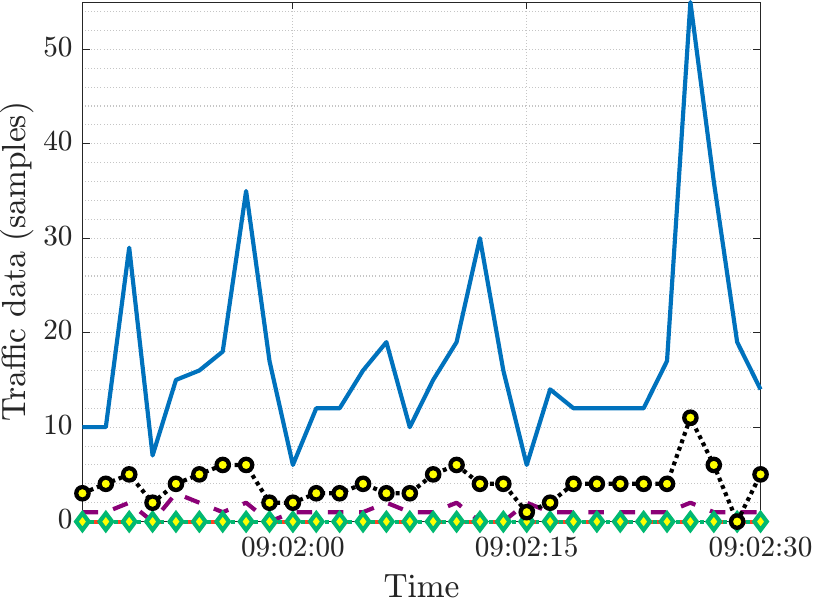}
        \caption{MitM attack state}
        \label{fig:realdec_5}
    \end{subfigure}
    \hfill
    \begin{subfigure}{0.3\linewidth}
        \centering
        \centerline{\includegraphics[width=.4\linewidth]{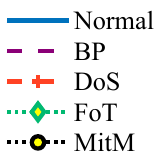}}
        \vspace*{1.5cm}
    \end{subfigure}
    \hfill
    \caption{Real-time blockchain cyberattack detection: proposed CoL model of 3 LNs in Ethereum node 1.}
    \label{fig:realdec}
    \vspace*{-0.5cm}
\end{figure*}

Fig.~\ref{fig:realdec} illustrates the real-time monitoring of our proposed CoL for normal state and three types of attacks in the network. Fig.~\ref{fig:realdec_1} is the normal state (Class-0) of the network with a high number of normal samples and a low number of attack samples. Then, the BP and MitM attacks are performed. Fig.~\ref{fig:realdec_2} and Fig.~\ref{fig:realdec_5} show a slight increase in the number of BP attacks and MitM attacks. 
This is because the number of BP attack samples is much smaller than other states in a period of time as in Table~\ref{tab:realsample}.
In this case, the detection mechanism is activated and alarms the network under the BP attack (Class-1). Similarly, the DDoS attack in the network is described in Fig.~\ref{fig:realdec_3} with a high increase in the number of samples of DoS attacks. Finally, Fig.~\ref{fig:realdec_4} describes the FoT attacks. Unlike other attacks, the FoT attacks include a large number of samples, thus it increases both the number of normal and attack samples (above 200~traffic~samples per 2~seconds) more than other attacks (about 100~traffic~samples per 2~seconds). Thereby, in all the cases, we can observe that our proposed intrusion detection system can detect attacks effectively in a real-time manner.    

\subsubsection{Real-time Processing Capacity}
\begin{figure}[t!]
	\centering
	\includegraphics[width=0.9\linewidth]{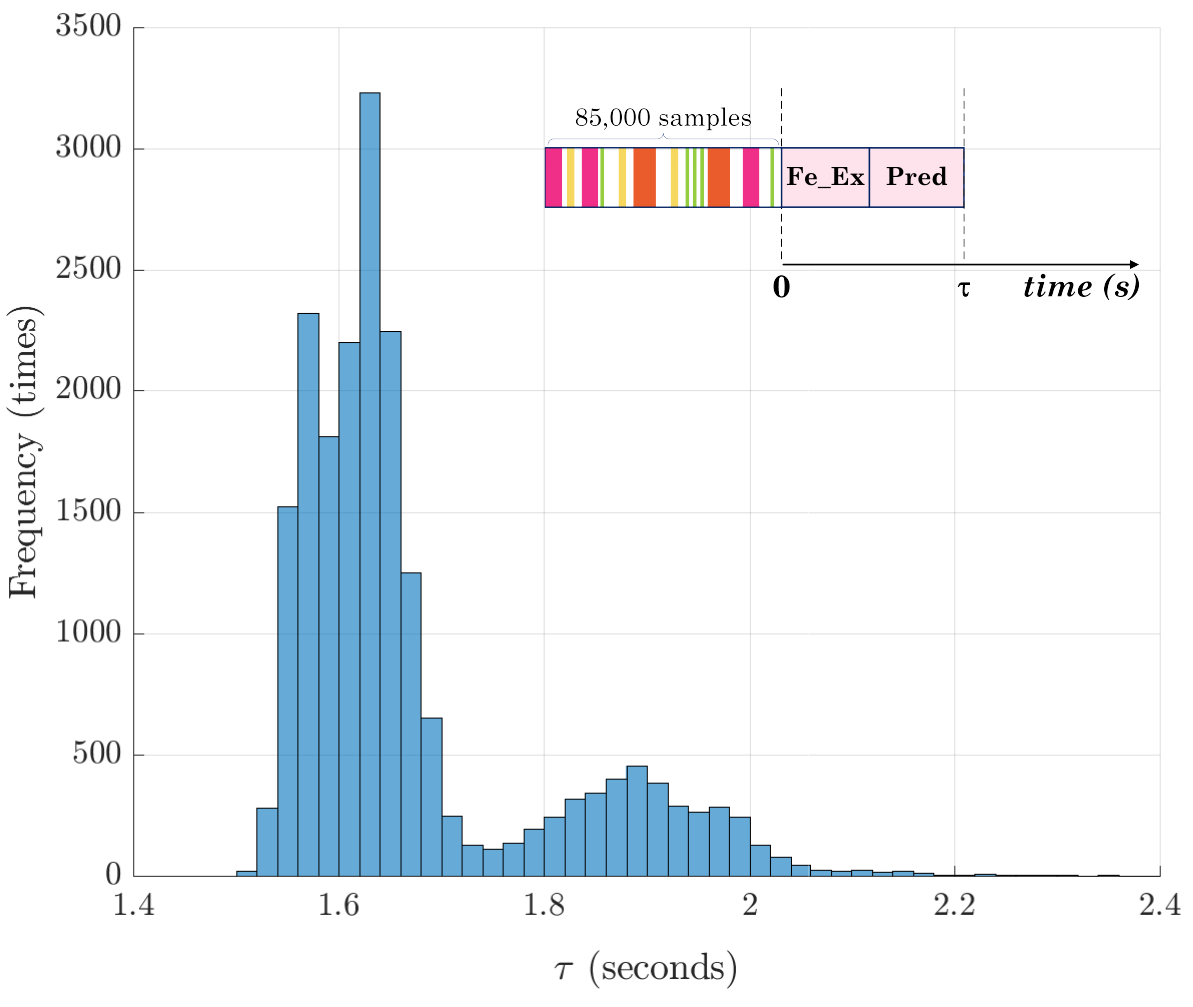}
	\caption{Histogram of real-time detection duration.}
	\label{fig:real_duration}
    \vspace*{-0.5cm}
\end{figure}

In this experiment, we fix the number of input samples in our proposed model to find the maximum real-time processing capacity in capturing and detecting attacks. Fig.~\ref{fig:real_duration} illustrates the real-time processing capacity of our proposed model. The processing time $\tau$ is counted from the time when our proposed model reads the file containing the samples, until completing classification and producing the output. This work is repeated 20,000 times to determine the stability of the detection time of our proposed model. We vary the number of input samples multiple times to find the appropriate number that is adapted to the condition in Fig.~\ref{fig:Real_exp_timeline}. In most of the cases (98\%), our proposed framework can classify 85,000 samples in less than 2~seconds. These results demonstrate that our proposed detection framework is efficient to deploy to detect attacks in real-world blockchain networks. It can not only detect attacks with high accuracy (up to 98.6\%) but also quickly (up to 85,000 samples within 2 seconds). 

\begin{table}[t]
\centering
\caption{The number of samples on LN-1 in five~hours.}
\label{tab:realsample}
\resizebox{0.49\textwidth}{!}{
\begin{tabular}{|l|c|c|c|c|c|}
\hline
\multicolumn{1}{|c|}{}           & \textbf{Class-0} & \textbf{Class-1} & \textbf{Class-2} & \textbf{Class-3} & \textbf{Class-4} \\ \hline
\textbf{Number of samples}       & 736,897           & 2,424              & 481,532            & 886,389 & 3,483          \\ \hline
\textbf{Portion (\%)} & 34.912           & 0.115            & 22.814            & 41.994  & 0.165         \\ \hline
\end{tabular}
}
\vspace*{-0.5cm}
\end{table}

\section{Conclusion}\label{sec:conclusion}

In this work, we have proposed a novel collaborative learning framework for a cyberattack detection system in a blockchain network. First, we have implemented a private blockchain network in our laboratory. This blockchain network is used to (1) generate data (both normal and attack data) to serve the proposed learning models and (2) validate the performance of our proposed learning framework in real-time experiments. After that, we have proposed a highly-effective learning model that allows to be effectively deployed in the blockchain network. This learning model allows nodes in the blockchain can be actively involved in the detection process by collecting data, learning knowledge from their data, and then exchanging knowledge together to improve the attack detection ability. In this way, we can not only avoid problems of conventional centralized learning (e.g., congestion and single point of failure) but also protect the blockchain network right at the edge. Both simulation and real-time experimental results then have clearly shown the efficiency of our proposed framework. In future, we plan to continue developing this dataset with other emerging types of attacks and develop more effective methods to protect blockchain networks.

\balance
\bibliographystyle{IEEEtran}
\bibliography{library}

\end{document}